\documentstyle[12pt,psfig]{article}
\begin{document}
\newcommand{\eett}{$e^+e^- \rightarrow t\overline{t}~$}
\newcommand{\ttbar}{$t\overline{t}~$}
\newcommand{\eebar}{$e^+e^-~$}
\def \beq{\begin{equation}}
\def \eeq{\end{equation}}
\def \bea{\begin{eqnarray}}
\def \eea{\end{eqnarray}}
\def \tt {t\overline {t}}
\def \ee {e^+e^-}
\def \ra {\rightarrow}
\def \g {\gamma}
\def \cvg {c_v^{\gamma}}
\def \cag {c_a^{\gamma}}
\def \cvz {c_v^Z}
\def \caz {c_a^Z}
\def \cdg {c_d^{\gamma}}
\def \cdz {c_d^Z}
\def \cdgz {c_d^{\gamma,Z}}
\def \t{$t\;$}
\def \toverline{$\overline t~$}
\def \tbar{\overline t}
\def \el{E_l}
\def \thetal{\theta_l}
\def \phil{\phi_l}
\def \f{\frac}
\def \o{\overline}
\def \half{\frac{1}{2}}

\thispagestyle{empty}
\begin{flushright}
PRL-TH-00/1\\
hep-ph/0002006
\end{flushright}

\begin{center}
{\boldmath 
\large \bf Effect of anomalous $tbW$ vertex on decay-lepton distributions in 
$e^+e^- \rightarrow t\overline{t}$ and \\ CP-violating asymmetries}
\end{center}

\vskip .5cm
\centerline{Saurabh D. Rindani}
\vskip .2cm
\centerline{\it Theory Group, Physical Research Laboratory}
\centerline{\it Navrangpura, Ahmedabad 380 009}
\vskip 1.5cm
\centerline{\bf Abstract}
\vskip .2cm

We obtain analytic expressions for the energy and polar-angle double
differential distributions of a secondary lepton $l^+$ ($l^-$) arising from the
decay of $t$ ($\tbar$) in \eett with an anomalous $tbW$ decay vertex. We also
obtain anlaytic expressions for the various differential cross sections with the
lepton energy integrated over. In this case, we find that the angular
distributions of the secondary lepton do not depend on the anomalous coupling
in the decay, regardless of possible anomalous couplings occurring in the
production amplitude for \eett. Our study includes the effect of longitudinal
$e^-$ and $e^+$ beam polarization. We also study the lepton energy and beam
polarization dependence of certain CP-violating lepton angular asymmetries
arising from an anomalous $tbW$ decay vertex and compare them with 
the asymmetries arising due to CP violation in the production process
due to the top electric or weak dipole moment. 

\newpage

\section{Introduction}
The standard model (SM) has been found to be in agreement with experiment 
in many of its
aspects. However, the properties of the top quark, in particular, the nature
and strength of its couplings, have yet to be studied accurately. A future
linear $\ee$ collider, operating at a centre-of-mass (cm) energy above the
$\tt$ threshold, will be able to determine with greater accuracy 
the couplings of the top quark to
the gauge bosons $\gamma$, $Z$, $W$, and possibly, to the Higgs \cite{desy}. 
A comparison of these with SM expectations will be able to shed light 
on new physics effects, if any.

The role of a linear $\ee$ collider (LC) in probing $\tt$ interactions
through the polarization of the $t$ and/or $\tbar$ has
received a fair amount of attention \cite{toppol, arens, parke}. 
In particular, the polarization effects
can be used to probe the anomalous magnetic and electric dipole couplings of
$t$, $\tbar$ to $\gamma$ and $Z$ [5-11],
as well as chromoelectric and chromomagnetic
dipole couplings to gluons \cite{michael}. A common underlying
idea behind these studies is, of course, that the top quark being heavy decays
before it can hadronize \cite{heavytop}. Hence its decay distributions would
retain information about the spin of the quark and 
would be useful in analyzing its polarization. The  
top quark polarization can thus be observed using its dominant
decay into a bottom quark and a $W$, the latter decaying into a quark pair or
a charged lepton and a neutrino. The accuracy of polarization measurements
is therefore dependent on an accurate  knowledge of the $tbW$ vertex. 
It is of utmost
importance, then, to study possible anomalous contributions to the $tbW$
vertex, which might arise from the same sources as anomalous contributions to
the production process.

In this paper we look at the process \eett , with either of $t$ or $\tbar$
decaying hadronically, while the other decays into a $b$ quark, a charged
lepton  and a neutrino. We will be interested only in the kinematic
distribution of the charged lepton as a probe of anomalous interactions either
in the production process (\eett ), or in the $tbW$ decay vertex. Our aim would
be to examine features of various distributions which might isolate or
emphasize anomalous effects in production over those in decay, or vice versa.

To this aim we have obtained analytic expressions for a full angular
distribution, and also combined energy and polar-angle distribution of the
secondary decay lepton in the cm frame, in the presence of non-standard
contributions to production as well as decay. We have also included the effect
of $e^+$ and $e^-$ beam longitudinal polarization. 

The energy distribution of a single lepton in a semileptonic process described
above, as well as energy correlations between leptons arising from $t$ and
$\tbar$ decay, have been studied before in \cite{asymm,PP,hiokie}. 
Angular distribution of
leptons, as well as angular correlations have been studied in the context of 
CP-violating (electric and weak dipole) moments of the top quark alone [5-11]
and/or of CP violation in decay \cite{decay}.
Many of these studies have been done in
the context of specific models. An analytic expression for combined lepton 
energy
and polar-angle distribution in the context of anomalous couplings present in
the top production as well as decay, which has been obtained here, did not
exist in the literature until very recently \cite{hiokilast}.
\footnote{The paper of Grzadkowski and Hioki \cite{hiokilast}, which 
appeared while this work was being completed, 
also calculates the double distribution mentioned above. 
However, the form they exhibit is less
explicit.} It should be noted, however, that our approach
is considerably different from that of \cite{hiokilast}. 
For example, we have used  helicity amplitudes rather
than the method of Kawasaki, Shirafuji and Tsai \cite{tsai} used by them to get
decay lepton distributions. 
We thus provide an independent cross check and confirmation of their 
expressions.

As it turns out, without the observation of lepton energy, it is impossible to
see the effect of anomalous $tbW$ vertex in the lepton angular distribution. It
is thus important to look at combined energy-angle distribution.

We have made use of our expressions to study CP violation in the production and
decay of $\tt$. We have calculated the contribution of the different sources of
CP violation, viz., top electric and weak dipole couplings and a CP-violating
$tbW$ vertex, to the dependence on lepton energy of certain simple angular
asymmetries.  We have also examined the role of longitudinal $e^+$ and $e^-$
beam polarization in discriminating among these different sources of CP
violation. 

One of the results of this work is that so long as the lepton energy is
integrated over, the distributions in terms of lepton and/or top angles are 
independent of non-standard effects in top decay, regardless of whether there
are any non-standard effects in production or not.
This has two consequences. Firstly, this implies that lepton angular
distributions can be safely used for analyzing top polarization without fear of
an error coming from possible non-standard effects in the $tbW$ coupling. Thus,
angular distributions can be used to study non-standard effects in top
production without contamination from non-standard effects in decay. On the
other hand, it is also evident that for studying non-standard effects in top
decay, one has necessarily to look at the energy dependence. We have found that
crucial changes of sign of certain asymmetries with polarization, and with
lepton 
energy, can be used to enhance the relative importance of the contributions of
CP violation coming from the dipole moments of the top quark, and from the
decay.

The plan of the rest of the paper is as follows. In the next section we set up
our formalism and list helicity amplitudes for $t$ and $\tbar$ decay in the
leptonic channel. In Section 3 we obtain energy integrated charged lepton 
angular distributions. In Section 4 we obtain the distribution for the charged
leptons in terms of energy and polar angle. Section 5 contains the results and
discussion.

\section{Formalism and helicity amplitudes}

We describe in this section the calculation of 
helicity amplitudes in
$\ee \ra \tt$ and the subsequent decay $t \ra b l^+ \nu_l\;
(\overline{t} \ra
\overline{b} l^- \overline{\nu_l})$.  We adopt the narrow-width
approximation for
$t$ and $\overline{t}$, as well as for $W^{\pm}$ produced in
$t,\;\overline{t}$
decay.

  We assume the top quark couplings to $\g$ and $Z$ to be
given by the vertex factor  $ie\Gamma_\mu^j$, where
\beq\label{dipole}
\Gamma_\mu^j\;=\;c_v^j\,\g_\mu\;+\;c_a^j\,\g_\mu\,\g_5\;+
\;\f{c_d^j}{2\,m_t}\,i\g_5\,
(p_t\,-\,p_{\overline{t}})_{\mu},\;\;j\;=\;\g,Z,
\eeq
with
\bea
\cvg&=&\f{2}{3},\:\;\;\cag\;=\;0, \nonumber \\
\cvz&=&\f {\left(\f{1}{4}-\f{2}{3} \,x_w\right)}
{\sqrt{x_w\,(1-x_w)}},
 \\
\caz&=&-\f{1}{4\sqrt{x_w\,(1-x_w)}}, \nonumber
\eea
and $x_w=sin^2\theta_w$, $\theta_w$ being the weak mixing
angle.
We have assumed in (1) that the only addition to the SM
couplings $c^{{\g},Z}_{v,a}$ are the $CP$-violating electric and weak
dipole
form factors, $e\cdg/m_t$ and $e\cdz/m_t$, which are
assumed small. We will call these electric dipole moment (edm) and weak dipole
moment (wdm) for short. Including additional anomalous couplings, viz., vector,
axial vector and magnetic dipole couplings is not problem, and our
deriviation of distributions would go through in that case. However, numerical
calculations in this paper are restricted to only CP-violating effects and
hence we do not include other form factors in eq. (\ref{dipole}). 
Use has also been made of the Dirac equation in
rewriting the usual dipole coupling
$\sigma_{\mu\nu}(p_t+p_{\overline{t}})^{\nu}\g_5$ as
$i\g_5(p_t-p_{\overline{t}})_{\mu}$, dropping small corrections
to the vector and axial-vector couplings.  

We write the contribution of a general $tbW$ vertex to $t$ and $\tbar$ decays
as 
\bea
\Gamma^{\mu}_{tbW}& =& -\f{g}{\sqrt{2}}V_{tb} \o{u}(p_b)
\left[\gamma^{\mu}(f_{1L}
P_L+f_{1R}P_R)\f{}{} \right. \nonumber \\
&& \left. - \f{i}{m_W} \sigma^{\mu\nu} (p_t - p_b)_{\nu}
 (f_{2L}P_L+f_{2R}P_R) \right] u(p_t),
\eea
\bea
\o{\Gamma}^{\mu}_{tbW}& =& -\f{g}{\sqrt{2}}V_{tb}^* \o{v}(p_{\o{t}})
\left[\gamma^{\mu}(\o{f}_{1L} P_L + \o{f}_{1R} P_R )\f{}{} \right. \nonumber \\
&& \left. -
\f{i}{m_W} \sigma^{\mu\nu} (p_{\o{t}} - p_{\o{b}})_{\nu} (\o{f}_{2L}P_L+\o{f}
_{2R}P_R) \right]
v(p_{\o{b}}),
\eea
where $P_{L,R} =\half (1\pm \gamma_5)$, and $V_{tb}$ is the 
Cabibbo-Kobayashi-Maskawa matrix element, which we take to be equal to one.
If CP is conserved, the form factors $f$ above obey the relations
\beq
f_{1L}=\o{f}_{1L};\;\; f_{1R}=\o{f}_{1R},
\eeq
and
\beq
f_{2L}=\o{f}_{2R};\;\; f_{2R}=\o{f}_{2L}.
\eeq
Like $c_d^{\gamma}$ and $c_d^Z$ above, we will also treat $f_{2L,R}$ and
$\o{f}_{2L,R}$ as small, and retain only terms linear in them.
For the form factors $f_{1L}$ and $\o{f}_{1L}$, we retain their SM values,
viz., $f_{1L} = \o{f}_{1L} = 1$. $f_{1R}$ and $\o{f}_{1R}$ do not contibute in
the limit of vanishing $b$ mass, which is used here. Also, $f_{2L}$ and
$\o{f}_{2R}$ drop out in this limit.

The helicity amplitudes for $\ee \ra \g^*,Z^* \ra \tt$ in the
centre-of-mass (cm) frame, including $\cdgz$ 
couplings, have been given in \cite{asymm} (see
also Kane {\it et al.}, ref. \cite{toppol}), so we do not repeat
them here.  

To calculate the decay helicity amplitudes, we use the standard Dirac gamma
matrix representations, and the following forms for Dirac spinors with definite
helicity.

For the spinors for $l$, $b$ and $\nu$, and their antiparticles, all of which
are assumed massless, we use the representations
\beq\label{uspinor}
u_-(P) = \sqrt{P^0}\left(
\begin{array}{c}
- \sin \left(\frac{\Theta}{2}\right) e^{-i\Phi} \\
  \cos\left(\frac{\Theta}{2}\right) \\
  \sin \left(\frac{\Theta}{2}\right) e^{-i\Phi} \\
- \cos\left(\frac{\Theta}{2}\right) 
\end{array} 
\right);\;
v_+(P) = \sqrt{P^0}\left(
\begin{array}{c}
  \sin \left(\frac{\Theta}{2}\right) e^{-i\Phi} \\
- \cos\left(\frac{\Theta}{2}\right) \\
- \sin \left(\frac{\Theta}{2}\right) e^{-i\Phi} \\
  \cos\left(\frac{\Theta}{2}\right) 
\end{array} 
\right),
\eeq
where the subscript denotes the sign of the helicity and $\Theta$ and $\Phi$ are
 the polar and azimuthal angles of
the momentum $P$ of the respective particle or antiparticle.

For the spinors corresponding to $t$ and $\o{t}$ in their respective rest 
frames, we use
\beq\label{tspinors}
u_{t+} = \sqrt{2m_t} \,\left(
\begin{array}{c}
1\\ 0\\ 0\\ 0 
\end{array} \right); \;
u_{t-} = \sqrt{2m_t} \,\left(
\begin{array}{c}
0\\ 1\\ 0\\ 0 
\end{array} \right), 
\eeq
\beq\label{tbarspinors}
v_{t+} = \sqrt{2m_t} \,\left(
\begin{array}{c}
0\\ 0\\ 1\\ 0 
\end{array} \right); 
v_{t-} = \sqrt{2m_t} \,\left(
\begin{array}{c}
0\\ 0\\ 0\\ 1 
\end{array} \right). 
\eeq

The non-vanishing
helicity amplitudes, respectively $M$ and $\overline{M}$, for
\[t \ra b W^+,\;
\;W^+ \ra l^+ \nu_l\]
and \[\overline{t} \ra \overline{b}W^-,\;\; W^- \ra
l^-\overline{\nu}_l\]
in the respective rest frames of $t$, $\overline{t}$, are given
below (we neglect all fermion masses except $m_t$, the top mass):
\bea
\lefteqn{M_{+-+-}=-2\sqrt{2}g^2\Delta_W(q^2)}\nonumber\\
&\times&\left\{ 
\left(1+\frac{f_{2R}}{\sqrt{r}}\right) \cos\f{\theta_{l^+}}
{2}\left[\cos \f{\theta_{\nu_l}}{2} \sin
\f{\theta_b} {2} e^{i\phi_b} - \sin \f{\theta_{\nu_l}}{2} 
\cos \f{\theta_b} {2} e^{i\phi_{\nu_l}}\right]\right.
 \nonumber\\
&&- \left.\f{f_{2R}}{\sqrt{2}}\sin{ \f{\theta_b}{2}}
e^{i\phi_b}
\left[\sin{\f{\theta_{\nu_l}}{2}}\sin{\f{\theta_{l^+}}{2}}
e^{i(\phi_{\nu_l}-\phi_{l^+})} + \cos{\f{\theta_{\nu_l}}{2}}
\cos{\f{\theta_{l^+}}{2}}\right]\right\}
,\nonumber \\
&& \\
\lefteqn{M_{--+-}=-2\sqrt{2}g^2\Delta_W(q^2)}\nonumber\\
&\times&\left\{ 
\left(1+\frac{f_{2R}}{\sqrt{r}}\right) \sin \f{\theta_{l^+}}
{2}e^{-i\phi_{l^+}}\left[\cos \f{\theta_{\nu_l}}{2} \sin
\f{\theta_b} {2} e^{i\phi_b} - \sin \f{\theta_{\nu_l}}{2} 
\cos \f{\theta_b} {2} e^{i\phi_{\nu_l}}\right]\right.
 \nonumber\\
&&+ \left.\f{f_{2R}}{\sqrt{r}}\cos{ \f{\theta_b}{2}}
\left[\sin{\f{\theta_{\nu_l}}{2}}\sin{\f{\theta_{l^+}}{2}}
e^{i(\phi_{\nu_l}-\phi_{l^+})} + \cos{\f{\theta_{\nu_l}}{2}}
\cos{\f{\theta_{l^+}}{2}}\right]\right\}
,\nonumber \\
&& \\
\lefteqn{\overline{M}_{++-+}=-2\sqrt{2}g^2\Delta_W(q^2)}
\nonumber\\
&\times&\left\{ 
\left(1+\frac{\overline{f}_{2L}}{\sqrt{r}}\right) 
\cos\f{\theta_{l^-}}
{2}\left[\cos \f{\theta_{\o{\nu_l}}}{2} \sin
\f{\theta_{\o{b}}} {2} e^{-i\phi_{\o{b}}} 
- \sin \f{\theta_{\o{\nu_l}}}{2} 
\cos \f{\theta_{\o{b}}} {2} e^{-i\phi_{\o{\nu_l}}}\right]
\right.
 \nonumber\\
&&- \left.\f{\o{f}_{2L}}{\sqrt{r}}\sin{ \f{\theta_{\o{b}}}{2}}
e^{-i\phi_{\o{b}}}
\left[\sin{\f{\theta_{\o{\nu_l}}}{2}}\sin{\f{\theta_{l^-}}{2}}
e^{i(\phi_{l^-}-\phi_{\o{\nu_l}} )}
+ \cos{\f{\theta_{\o{\nu_l}}}{2}}
\cos{\f{\theta_{l^-}}{2}}\right]\right\}
,\nonumber \\
\lefteqn{\overline{M}_{-+-+}=-2\sqrt{2}g^2\Delta_W(q^2)}
\nonumber\\
&\times&\left\{ 
\left(1+\frac{\overline{f}_{2L}}{\sqrt{r}}\right) 
\sin\f{\theta_{l^-}}{2}e^{i\phi_{l^-}}
\left[\cos \f{\theta_{\o{\nu_l}}}{2} \sin
\f{\theta_{\o{b}}} {2} e^{-i\phi_{\o{b}}} 
- \sin \f{\theta_{\o{\nu_l}}}{2} 
\cos \f{\theta_{\o{b}}} {2} e^{-i\phi_{\o{\nu_l}}}\right]
\right.
 \nonumber\\
&&+ \left.\f{\o{f}_{2L}}{\sqrt{r}}\cos{ \f{\theta_{\o{b}}}{2}}
\left[\sin{\f{\theta_{\o{\nu_l}}}{2}}\sin{\f{\theta_{l^-}}{2}}
e^{i(\phi_{l^-}-\phi_{\o{\nu_l}}) }
+ \cos{\f{\theta_{\o{\nu_l}}}{2}}
\cos{\f{\theta_{l^-}}{2}}\right]\right\},
\eea
where
\beq
\Delta_W(q^2)=\f{1}{q^2-m_W^2+i\Gamma_Wm_W}
\eeq
is the $W$ propagator, $q$ its momentum, and
\beq
r=\f{m_W^2}{m_t^2}.
\eeq
The subscripts $\pm$ refer to signs of the helicities, the order of
the helicities being $t$, $b$, $l^+$, $\nu_l$ ($\overline{t}$,
$\overline{b}$, $l^-$, $\overline{\nu_l}$).
The various $\theta$'s are polar angles of the particles (in the $t$ rest
frame) and of the 
antiparticles (in the $\overline{t}$ rest frame) labelled by
the suffixes with respect to a $z$ axis which is the direction in 
which the
top momentum is boosted to go from its rest frame to the cm frame. $\phi$'s
are the azimuthal angles with respect to an $x$ axis chosen in the
plane containing the $e^-$ momentum and positive $z$ directions.
We could have chosen here the polar and azimuthal angles of the lepton to be
zero in the top rest frame. However, we maintain nonzero values for future use,
when we boost to the cm frame.

The density matrix elements calculated from the above 
helicity amplitudes in the rest frame of the top are given
by
\beq
\Gamma_{\pm\pm} = g^4\vert\Delta(q^2)
\vert^2 (m_t^2-2p_t\cdot p_{l^+})(1\pm \cos{\theta_{l^+}})
\left( 1+ \f {{\rm Re}f_{2R}}{\sqrt{r}}\f{m_W^2}{p_t\cdot p_{l^+}}
\right),
\eeq
\beq
\Gamma_{\pm\mp} = g^4\vert\Delta(q^2)
\vert^2 (m_t^2-2p_t\cdot p_{l^+})\sin{\theta_{l^+}}e^{\pm i
\phi_{l^+}}
\left( 1+ \f {{\rm Re}f_{2R}}{\sqrt{r}}\f{m_W^2}{p_t\cdot p_{l^+}}
\right),
\eeq
\beq
\o{\Gamma}_{\pm\pm} = g^4\vert\Delta(q^2)
\vert^2 (m_t^2-2p_{\o{t}}\cdot p_{l^-})(1\pm \cos{\theta_{l^-}})
\left( 1+ \f {{\rm Re}\o{f}_{2L}}{\sqrt{r}}\f{m_W^2}{p_{\o{t}}\cdot p_{l^-}}
\right),
\eeq
\beq
\o{\Gamma}_{\pm\mp} = g^4\vert\Delta(q^2)
\vert^2 (m_t^2-2p_{\o{t}}\cdot p_{l^-})\sin{\theta_{l^-}}e^{\mp i
\phi_{l^-}}
\left( 1+ \f{{\rm Re}\o{f}_{2L}}{\sqrt{r}}\f{m_W^2}{p_{\o{t}}\cdot p_{l^-}}
\right),
\eeq
where an averaging over the the azimuthal angles $\phi_b$ and $\phi_{\o{b}}$ has
been done.
Notice that the only change in the decay density matrix relative to the
expression in SM is the overall factor $1+ \f {{\rm
Re}f_{2R}}{\sqrt{r}}\f{m_W^2}{p_t\cdot p_{l^+}}$ or $1+ \f {{\rm
Re}\o{f}_{2L}}{\sqrt{r}}\f{m_W^2}{p_{\o{t}}\cdot p_{l^-}}$.
This has the important consequence that regardless of
any anomalous contributions to the production process, the decay-lepton 
double differential
distribution (calculated below) gets modified by anomalous $tbW$ couplings in
the narrow-width approximation by the same overall factor. It is also
interesting to note that only the real part of the anomalous couplings appear
in the overall factor. This actually corresponds to the absorptive part of the
decay amplitude.

We now make a Lorentz transformation to the laboratory frame using the fact 
that $\Gamma_{ij}$ and $\o{\Gamma}_{ij}$ are invariant, and the following 
transformations for the angles:
\beq
1\pm \cos{\theta_{l^+}} = \f{(1\mp \beta )(1\pm \cos{\theta^{\rm c.m.}_{tl^+}})}
{1-\beta \cos{\theta^{\rm c.m.}_{tl^+}}},
\eeq
\bea
\sin{\theta_{l^+}}e^{i\phi_{l^+}}&=&\f{\sqrt{1-\beta^2}}{1 
- \beta \cos{\theta^{\rm c.m.}
_{tl^+}}}
\left(\sin{\theta_{l^+}^{c.m.}}\cos{\theta_t^{c.m.}}\cos{\phi_{l^+}^{c.m.}}
\right. \nonumber\\
&& \left. 
-\cos{\theta_{l^+}^{c.m.}}\sin{\theta_t^{c.m.}}
+ i \sin{\theta_{l^+}^{c.m.}}\sin{\phi_{l^+}^{c.m.}}\right).
\eea
where $\theta^{\rm c.m.}_{tl^+}$ is the angle between $t$ and $l^+$ directions
in the cm frame.
There are similar expressions for the angles of $\o{t}$, and its decay
products, which we do not give explicitly. We will henceforth calculate 
everything in the cm frame. 
We will therefore drop the label ``c.m."
over the angles, it being understood that all variables are in the $e^+e^-$ c.m.
frame.

Combining the
production and decay density matrices in the narrow-width
approximation for $t,\overline{t},W^+,W^-$,
we get the $l^+$ and $l^-$ distributions for the case of $e^-$, $e^+$ with
polarization $P_e$, $P_{\o e}$ to be:
\begin{eqnarray}\label{dist}
\lefteqn{\frac{d\sigma^{\pm}}{d\cos\theta_t dE_l d\cos\theta_l
d\phi_l}= \frac{3\alpha^4\beta}{16x_w^2\sqrt{s}}
 \frac{E_l}{\Gamma_t \Gamma_W m_W}}
\nonumber \\
&\times&
\left( 1 + \f{{\rm Re}f^{\pm}}{\sqrt{r}}\f{2 m_W^2}{E_l\sqrt{s}
(1 - \beta \cos{\theta_{tl}})} \right)
\left(  \frac{1}{1-\beta\cos\theta_{tl}}
  -\frac{4 E_l
}{\sqrt{s}(1-\beta^2)}\right)\nonumber \\
&\times & \left\{ A (1-\beta \cos \theta_{tl})
+ B^{\pm} (\cos\theta_{tl}-\beta )\right. \nonumber \\
&&+\left. C^{\pm} 
(1-\beta^2) \sin\theta_t \sin\theta_l (\cos\theta_t \cos\phi_l -
\sin\theta_t \cot\theta_l)\right. \nonumber \\
&&+ \left.   D^{\pm}
(1-\beta^2) \sin\theta_t \sin\theta_l \sin\phi_l \right\},
\end{eqnarray}
where $A$, $B$, $C$ and $D$ 
are quantities related to the production density matrices $\rho_{ij}$ and 
$\overline{\rho}_{ij}$ for $t$ and $\o{t}$ respectively, by
\beq
A=\rho_{++} + \rho_{--}=\o{\rho}_{++} + \o{\rho}_{--},
\eeq
\beq
B^+=\rho_{++} - \rho_{--},
\eeq
\beq
B^-=\o{\rho}_{++} - \o{\rho}_{--},
\eeq
\beq
C^+={\rm Re}\rho_{+-}\f{\sqrt{s}}{m_t\sin\theta_t},
\eeq
\beq
C^-={\rm Re}\o{\rho}_{+-}\f{\sqrt{s}}{m_t\sin\theta_t},
\eeq
\beq
D^+={\rm Im}\rho_{+-}\f{\sqrt{s}}{m_t\sin\theta_t},
\eeq
\beq
D^-={\rm Im}\o{\rho}_{+-}\f{\sqrt{s}}{m_t\sin\theta_t}.
\eeq

The quantities $A_i$, $B_i$, $C_i$ and $D_i$ occurring in the above
equations are functions of the masses, $s$, the degrees of $e^-$ and 
${e^+}$  polarization ($P_{e^-}$ and $P_{e^+}$), and the coupling constants.
They are listed in the appendix of \cite{ppdist}.

In eq. (\ref{dist}), $\sigma^+$ and $\sigma^-$ refer respectively to $l^+$ and
$l^-$ distributions, with the same notation for the kinematic
variables of
particles and antiparticles.  Thus, $\theta_t$ is the polar angle of
$t$  (or $\tbar$), and $\el,\;\thetal,\;\phil$ are the energy, polar
angle and azimuthal angle of $l^+$ (or $l^-$).
All the angles are now in the cm frame, with the $z$ axis chosen
along the $e^-$ momentum, and the $x$ axis chosen in the plane
containing the $e^-$ and $t$ directions.
$\theta_{tl}$ is the angle between
the $t$ and $l^+$ directions (or $\overline t$ and $l^-$ directions).
$\beta$
is the $t$ (or $\o t$) velocity: \(\beta=\sqrt{1-4m_t^2/s}\).
Also, we use $f^+$ and $f^-$ to denote $f_{2R}$ and $\o{f}_{2L}$, 
respectively. We note that the only effect of the anomalous decay vertex is
to multiply the differential cross section by an overall factor, which is 
dependent on the lepton energy and the decay angle of the lepton with respect
to the top direction. 

We can now proceed in either of two ways. If we are not interested in energy
dependence, we can integrate eq. (\ref{dist}) over
$E_l$, and then over $\phi_l$, and finally over $\cos\theta_t$, as done in ref.
\cite{ppdist} to get the $d\sigma /d\cos\theta_l$. In the intermediate steps,
we would get angular distributions which contain dependence on the top-quark
polar angle. We follow this procedure in the next
section. Alternatively, if we are interested in the double differential cross
section $d^2\sigma /(dE_ld\cos\theta_l)$, we have to proceed somewhat
differently. That procedure is outlined in Sec. 4.

\section{Angular distributions}
We now proceed to obtain an expression for angular distributions as outlined
earlier. We start with eq. (\ref{dist}) and carry out the $E_l$ integration.
The limits of integration are
\beq\label{elimits}
\frac{m_W^2}{\sqrt{s}}\f{1}{1-\beta\cos\theta_{tl}} \leq E_l \leq
\frac{m_t^2}{\sqrt{s}}\f{1}{1-\beta\cos\theta_{tl}}.
\eeq
The integration is simple to carry out, and we get the result
\bea\label{ctcldist}
\lefteqn{\frac{d\sigma^{\pm}}{d\cos\theta_t d\cos\theta_l
d\phi_l}= \frac{3\alpha^4\beta}{16x_w^2\sqrt{s}}
 \frac{1}{\Gamma_t \Gamma_W m_W}}\nonumber \\
&\times &\f{m_t^4}{6s}\left(1-r \right)^2
\left(1+2r -6 {\rm Re}f^{\pm}\sqrt{r}\right)
\f{1}{(1-\beta\cos\theta_{tl})^3}
\nonumber \\
&\times & \left\{ A (1-\beta \cos \theta_{tl})
+ B^{\pm} (\cos\theta_{tl}-\beta )\right. \nonumber
\\
&&+\left.   C^{\pm}
(1-\beta^2) \sin\theta_t \sin\theta_l (\cos\theta_t \cos\phi_l -
\sin\theta_t \cot\theta_l)\right. \nonumber \\
&&+ \left.   D^{\pm}
(1-\beta^2) \sin\theta_t \sin\theta_l \sin\phi_l \right\},
\end{eqnarray}
As this equation shows, the effect of anomalous $tbW$ couplings $f^{\pm}$ on
the distribution is through an overall constant factor $1+2r -6 {\rm
Re}f^{\pm}\sqrt{r}$. Any further integration will not affect this
factor. In fact, this is precisely the factor which enters the $t$ and $\o{t}$
total widths to first order in $f^{\pm}$:
\beq\label{gammat}
\Gamma_t = \f{\alpha^2}{24x_w^2}\f{m_t^3}{\Gamma_Wm_W}(1-r)^2[(1+2 r)
-6{\rm Re} f_{2R}\sqrt{r}], 
\eeq
with a similar expression for $\Gamma_{\o{t}}$, with $f_{2R}$ replaced by
$\o{f}_{2L}$. Using the expression for $\Gamma_t$ from eq. (\ref{gammat}) in 
eq. (\ref{ctcldist}), we get an expression which is identical to the one
obtained with SM couplings for $t$ and $\o{t}$ decay.

Thus, at least in the linear approximation scheme for anomalous decay couplings
which 
we are employing, anomalous
couplings (CP conserving as well as CP violating) in the decay have no effect
on the angular distributions. This holds for arbitrary longitudinal
polarizations of $e^+$ and $e^-$ beams and arbitrary anomalous $\gamma t\o{t}$
and $Zt\o{t}$ couplings, whether CP violating or CP conserving.
It should be clarified that this result does not depend on a linear 
approximation in
the anomalous $\gamma t\o{t}$ and $Zt\o{t}$ couplings, for then the result
would be trivial. The result holds for arbitrary values of the 
quantities $A$, $B$, $C$, $D$ parametrizing the
production density matrix. Thus these could include higher orders of dipole
couplings, or other anomalous couplings without affecting our result.

We can further do an integration over $\phi_l$ and $\cos\theta_t$ to obtain for
the single differential cross section the same expression as with SM couplings
for the $tbW$ vertex. This expression was given in \cite{ppdist}, and has
subsequently been found to agree with the result of Grzadkowski and Hioki by
them \cite{hiokilast}. However, in
view of typographical errors in the expression in \cite{ppdist}, we give the
correct form here:
\bea\label{cldist}
\lefteqn{
\f{d\sigma^{\pm}}{d\cos\theta_l}=\frac{3\pi\alpha^2}{32s}\beta
\left\{4A_0
\mp 2A_1\left(\f{1-\beta^2}{\beta^2} \log\f{1+\beta}{1-\beta}-
\f{2}{\beta}\right) \cos\theta_l \right.}\nonumber \\
&&\left. + 2A_2 \left(
\f{1-\beta^2}{\beta^3}\log\f{1+\beta}{1-\beta}
(1-3\cos^2\theta_l) \right. \right. \nonumber \\
&& \left. \left. - \f{2}{\beta^2} (1-3\cos^2\theta_l-\beta^2+2 \beta^2
\cos^2\theta_l) \right) \right. \nonumber \\
&&\left. \pm 2B_1\f{1-\beta^2}{\beta^2}  \left(
\f{1}{\beta}\log\f{1+\beta}{1-\beta} -
2 \right) \cos\theta_l \right. \nonumber \\
&&  \left. + B_2^{\pm} \f{1-\beta^2}{\beta^3}
\left( \f{\beta^2-3}{\beta} \log\f{1+\beta}{1-\beta} + 6 \right)
(1-3\cos^2\theta_l) \right. \nonumber \\
&&\left. \pm 
2C_0^{\pm}\f{1-\beta^2}{\beta^2} \left( \f{1-\beta^2}{\beta}
\log\f{1+\beta}{1-\beta} - 2 \right) \cos\theta_l
\right. \nonumber \\
&& \left. - C_1^{\pm}\f{1-\beta^2}{\beta^3} \left( \f{3(1-\beta^2)}{\beta}
\log\f{1+\beta}{1-\beta} -2(3-2\beta^2)\right) (1-3
\cos^2\theta_l) \right\},
\eea
The quantities $A_i$, $B_i$ and $C_i$ in the above equation are coefficients of
powers of $\cos\theta_t$, as defined below ($D_i$ do not appear in the above
equation, but we define them here for completeness):
\beq
A=A_0+A_1 \cos\theta_t + A_2 \cos^2\theta_t,
\eeq
\beq
B^{\pm} = B_0^{\pm} + B_1 \cos\theta_t + B_2^{\pm},
\cos^2\theta_t,
\eeq
\beq
C^{\pm}=C_0^{\pm} + C_1^{\pm} \cos\theta_t,
\eeq
\beq
D^{\pm}=D_0^{\pm} + D_1^{\pm} \cos\theta_t.
\eeq
The values of the coefficients are given in the appendix of \cite{ppdist}, in
the presence of CP-violating electric and weak dipole moments. We only note
here that contribution of CP-even magnetic and weak magnetic dipole moments may
easily be included in those expressions using the helicity amplitudes given,
e.g., by Ladinsky and Yuan in \cite{toppol}. However, we will not make use of
these in this paper. 

\section{Energy and angle double differential distributions}

In order to obtain a distribution only in terms of lepton variables, we need to
integrate the expression in (\ref{dist}) for the differential cross section over
$\cos\theta_t$. However, if we are interested only in the lepton 
energy and polar angle distributions, and not in the azimuthal angle of the the
lepton, it is more convenient to proceed as follows. 
We first make a change of variables
from $\cos\theta_t$ and $\phi_l$ to $\cos\theta_{tl}$ and $\alpha$, where
$\alpha$ is defined by
\beq
\cos\theta_t = \cos\theta_{tl}\cos\theta_l + \sin\theta_{tl}\sin\theta_l
\cos\alpha.
\eeq
We then have
\beq
d\cos\theta_t\,d\phi_l = d\cos\theta_{tl}\,d\alpha,
\eeq
and the relations for the sine of the angle $\theta_t$,
\beq
\sin\theta_t\sin\phi_l = \sin\theta_{tl}\sin\alpha ,\;
\sin\theta_t\cos\phi_l = \cos\theta_{tl}\sin\theta_l - \sin\theta_{tl}
\cos\theta_l\cos\alpha.
\eeq

We can now integrate (\ref{dist}) over $\alpha$ over the range 
$0$ to $2 \pi$ to get
\bea
\lefteqn{\f{d\sigma^{\pm}}{dE_l\,dc_l\,dc_{tl}} 
= \f{3\pi\alpha^4\beta}{32\sqrt{s}
\sin\theta_W^4}\f{E_l}{\Gamma_t\Gamma_Wm_W}}\nonumber\\
&\times&\left\{\left[ \left(A_0 - \f{B_0^{\pm}}{\beta}\right)
\pm \left(A_1-\f{B_1} {\beta}\right) c_l c_{tl} 
+ \left(A_2-\frac{B_2^{\pm}}{\beta}\right)(c_l^2 c_{tl}^2
+ \half s_l^2 s_{tl}^2)\right] \right.\nonumber\\
&  &+ \left. \left[ B_0^{\pm} \pm B_1 c_l c_{tl} 
+ B_2^{\pm} (c_l^2 c_{tl}^2 + \half s_l^2 s_{tl}^2) \right] 
\f{(1-\beta^2)}{\beta}\f{1}{(1-\beta c_{tl})} \right.\nonumber
\\
&&+ \left. \left[ \mp C_0^{\pm} c_l s_{tl}^2 
+ \half C_1^{\pm} (1-3 c_l^2) c_{tl} s_{tl}^2 
\right] \right\}\nonumber \\
&\times&\left[1 - \f{E_l \sqrt{s}}{m_t^2}(1-\beta c_{tl})\right]\left[1+ 
\f{2 {\rm Re}f^{\pm}}{ \sqrt{r}} \f{m_W^2}{E_l\sqrt{s}(1-\beta c_{tl})}\right].
\eea
In the above equation we have used the notation
\beq
c_l\equiv \cos\theta_l,\, s_l\equiv \sin\theta_l; \;\;
c_{tl}\equiv \cos\theta_{tl},\, s_{tl}\equiv \sin\theta_{tl}.
\eeq
Note that in the above equation, as well as in following equations, $\Gamma_t$
is given by eq. (\ref{gammat}) above, which includes the correction from the
anomalous decay vertex.

The $c_{tl}$ integrals can be carried out analytically, the limits of
integration depending on the lepton energy. These may be written as
\beq
L<c_{tl}<U,
\eeq
where $L$ and $U$ take different values in different energy ranges. They are 
given by
\beq
L = {\rm max.} \left( -1, \f{1}{\beta} \left( 1- \f{m_t^2}{E_l\sqrt{s}}\right)
	\right),\;
U = {\rm min.} \left( 1, \f{1}{\beta} \left( 1- \f{m_W^2}{E_l\sqrt{s}}\right)
       \right).
\eeq
For a given value of $\sqrt{s}$, this gives three possible ranges of energy,
with distinct sets of limits on $c_{tl}$. These ranges of energy are
described in detail in \cite{arens}.

The result of integration over $c_{tl}$ from $L$ to $U$ is given by:
\bea\label{double}
\lefteqn{\f{d\sigma^{\pm}}{dE_l\,dc_l} = \f{3\pi\alpha^4\beta}{32\sqrt{s}
\sin\theta_W^4}\f{E_l}{\Gamma_t\Gamma_Wm_W}}\nonumber\\
&\times& \left[ \left( A_0 -\f{B_0^{\pm}}{\beta}\right) X_0
\pm c_l\left( A_1 -\f{B_1}{\beta}\right) X_1\right.\nonumber\\
&&+ \left. \half (3c_l^2-1)\left( A_2 -\f{B_2^{\pm}}{\beta}\right) X_2
+ \half s_l^2\left( A_2 -\f{B_2^{\pm}}{\beta}\right) X_2'\right.\nonumber\\
&&+\left. \f{1-\beta^2}{\beta} \left[ \left( B_0^{\pm}+\half s_l^2 B_2^{\pm}
\right) Y_0 \pm B_1 c_l Y_1
+ \half B_2^{\pm} (3c_l^2-1) Y_2 \right]\right.\nonumber\\
&&\mp \left. (1-\beta^2) C_0^{\pm} c_l Z_0 
- \half (1-\beta^2) C_1^{\pm} (3c_l^2-1) Z_1 \right].
\eea

Here we have used
\beq
X_0 = P(U-L) +\half Q (U^2 -L^2) + \f{R}{\beta} {\cal L},
\eeq
\beq
X_1 = \half P(U^2-L^2) + \f{1}{3} Q(U^3-L^3) - \f{R}{\beta}F,
\eeq
\beq
X_2 = \f{1}{3}P(U^3-L^3)+\f{1}{4}Q(U^4-L^4) - \f{R}{\beta^2}
\left( F + \half \beta (U^2-L^2)\right) ,
\eeq
\beq
X_2' = P(U-L) + \half Q(U^2-L^2)+\f{R}{\beta}{\cal L},
\eeq
\beq
Y_0 = \f{1}{\beta}P' {\cal L} - \f{1}{\beta} Q (U-L)
 + R \f{U-L}{(1-L \beta)(1-U\beta)},
\eeq
\beq
Y_1 = - \f{1}{\beta}P' F -\half \f{Q}{\beta} (U^2-L^2)
+\f{R}{\beta}\left(\f{U-L}{(1-L \beta)(1-U\beta)} -\f{1}{\beta} {\cal L}\right),
\eeq
\bea
Y_2& =& -\f{1}{\beta^2}P' \left( F + \half \beta (U^2-L^2)
\right) -\f{Q}{3\beta}(U^3-L^3) \nonumber \\
& &+ \f{R}{\beta^2} \left( U-L -\frac{2}{\beta} {\cal L}
+ \f{U-L}{(1-L \beta )(1-U\beta )}\right),
\eea
\bea
Z_0& =& \f{1}{\beta^2}P' \left( U-L + \half \beta (U^2-L^2)
-\frac{1-\beta^2}{\beta} {\cal L}\right)\nonumber\\
& &+ \f{Q}{\beta}
\left( -(U-L) + \f{1}{3}(U^3-L^3)\right) \nonumber \\
&&+ \f{R}{\beta^2}\left(-2F + (U-L)
- \frac{(1-\beta^2)(U-L)}{(1-L\beta)(1-U\beta)}\right),
\eea
\bea
Z_1& =& \f{1}{\beta^3}P' \left((1-\beta^2)F + \half \beta
(U^2-L^2) +\f{\beta^2}{3}(U^3-L^3) \right)\nonumber \\
&&-\f{Q}{2\beta}\left( (U^2-L^2) - \f{1}{2}(U^4-L^4) \right) 
\nonumber \\
&&+ \f{R}{\beta^3}\left(- 2 F-\f{\beta}{2}
(U^2-L^2) + \f{1-\beta^2}{\beta} {\cal L} -
\frac{(1-\beta^2)(U-L)}{(1-L\beta)(1-U\beta)}\right) .
\eea
Also,
\beq
P= \left(1-\f{E_l \sqrt{s}}{m_t^2}\right)-2f^{\pm}\sqrt{r},
\eeq
\beq
Q=\f{E_l \sqrt{s}}{m_t^2}\beta ,
\eeq
\beq
R=\frac{m_W^2}{E_l\sqrt{s}}\frac{2 f^{\pm}}{\sqrt{r}},
\eeq
\beq
P'=P+Q/\beta =1-2f^{\pm}\sqrt{r},
\eeq
\beq
F = U-L - \f{1}{\beta} \log{\f{1-L\beta}{1-U\beta}},
\eeq
\beq
{\cal L} = \log{\f{1-L\beta}{1-U\beta}}.
\eeq

We can use the above expression for the double differential cross section to
obtain CP-violating asymmetries. For example, the diference in the differential
cross sections for $l^+$ and $l^-$ for a given value of $E_l$ but for values
of $\cos\theta_l$ differing in sign is a CP-odd quantity. The double
differential cross sections can easily be integrated to obtain analytic
expressions for certain angular asymmetries as a function of the lepton energy.
We had earlier proposed two asymmetries in the case where the lepton energy was
completed integrated over \cite{ppdist,ppasymm}, viz. the charge
asymmetry (with a cut-off $\theta_0$ in forward and backward angles), 
and the sum of the
forward-backward asymmetries (again with an angular cut-off) for $l^+$ and
for $l^-$.

We redefine these for the case when energy is not integrated over:  
\beq
{\cal A}_{ch}(E_l,\theta_0)=\frac{
{\displaystyle          \int_{\theta_0}^{\pi-\theta_0}}d\theta_l
{\displaystyle          \left( \frac{d\sigma^+}{dE_ld\theta_l}
                -       \frac{d\sigma^-}{dE_ld\theta_l}\right)}}
{
{\displaystyle          \int_{\theta_0}^{\pi-\theta_0}}d\theta_l
{\displaystyle          \left( \frac{d\sigma^+}{dE_ld\theta_l} +
\frac{d\sigma^-}{dE_ld\theta_l}\right)}},
\eeq
\beq
{\cal A}_{FB}(E_l,\theta_0)= \frac{ {\displaystyle
\int_{\theta_0}^{\frac{\pi}{2}}}d\theta_l {\displaystyle
\left( \frac{d\sigma^+}{dE_ld\theta_l} +
\frac{d\sigma^-}{dE_ld\theta_l}\right)} {\displaystyle
-\int^{\pi-\theta_0}_{\frac{\pi}{2}}}d\theta_l {\displaystyle
\left( \frac{d\sigma^+}{dE_ld\theta_l} +    \frac{d\sigma^-}{dE_ld\theta_l}
\right)}}
{
{\displaystyle          \int_{\theta_0}^{\pi-\theta_0}}d\theta_l
{\displaystyle          \left( \frac{d\sigma^+}{dE_ld\theta_l} +
\frac{d\sigma^-}{dE_ld\theta_l}\right)}}.
\eeq
The first of these, for the case when there is no angular cut-off, and with
only CP violation in top production included, was discussed in \cite{asymm, 
easy} for unpolarized beams, and in \cite{PP, 
easypol} for longitudinally polarized beams.

We will calculate these asymmetries as functions of the lepton energy. 

We do not display the expressions for these asymmetries. The corresponding
angular integration is trivial, and the expressions are lengthy. 

In the next section we apply the above expressions to obtain numerical values
for distributions and asymmetries.

\section{Results and Discussion}

We will use expressions obtained in the previous sections to look at
CP-violating asymmetries arising from top edm, wdm, and from $tbW$ vertex. The
asymmetries would get a contribution simultaneously from all these three
sources. However, we have treated these sources one at a time, with the
understanding that contributions from these would get added linearly in the
asymmetries. In a given model, these contributions would occur in a fixed
linear combination. However, in  a model-independent approach, methods have to
be devised to obtain the parameters simultaneously, independent of one another.

In our calculations in this section, we shall assume an LC operating at a cm
energy of 500 GeV. We will assume a top-quark mass of 174
GeV and $x_w=0.23$.
We assume that while one of $t$ and $\tbar$ decays into
a leptonic channel containing either $e$ or $\mu$, the other decays
hadronically into $b$ and two jets. This means that all the earlier expressions
for differential cross sections have to be multiplied by the product of the
hadronic branching ratio 2/3 and the leptonic branching ratio 2/9.

We have assumed that electron and positron beams are unpolarized, or fully
longitudinally polarized, with the electron and positron beam polarizations
equal and opposite. This may clearly not be possible in practice, since
polarization of positron beams is more difficult. 
However, this strong assumption is not necessary, since
what appears in the expressions for asymmetries is only the effective
polarization 
\beq 
P= (P_{e^-}-P_{e^+})/(1 - P_{e^-}P_{e^+}).
\eeq
Thus, for example, it
is sufficient if $P_{e^-}=-1$ and $P_{e^+}=0$ for $P$ to be $-1$.\footnote{
When $P_{e^-}\neq
-P_{e^+}$, the initial state is not really a CP eigenstate. However, for
practical purposes, this does not lead to any problem \cite{back}.}

 The
expressions for absolute (differential) cross sections would depend on $P$ as
well as on $1 - P_{e^-}P_{e^+}$. In what follows, we will label polarizations
by $P$.

\begin{figure}[tpb]
\begin{center}
\vskip -1cm
\centerline{\protect\hbox{\psfig{file=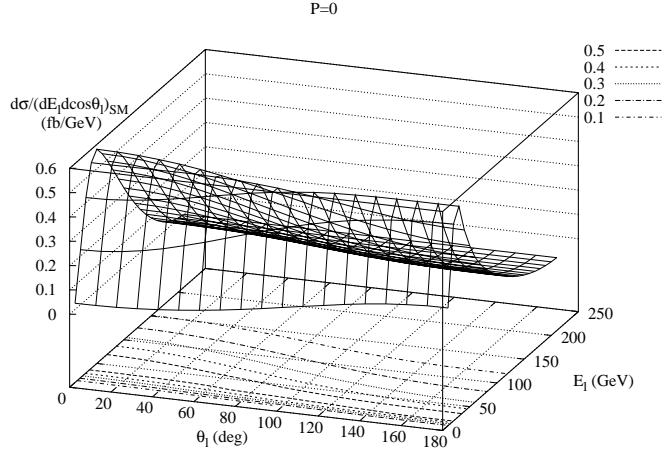,height=7cm}}}
\caption{The standard model double differential cross section as a function of
energy $E_l$ and polar angle $\theta_l$ of the charged lepton $l^+$ (either
$\mu^+$ or $e^+$) for unpolarized beams. The bottom plane shows a few contours
of constant cross section.}
\label{distpe0}
\end{center}
\end{figure}

\begin{figure}[tpb]
\begin{center}
\vskip -2cm
\centerline{\protect\hbox{\psfig{file=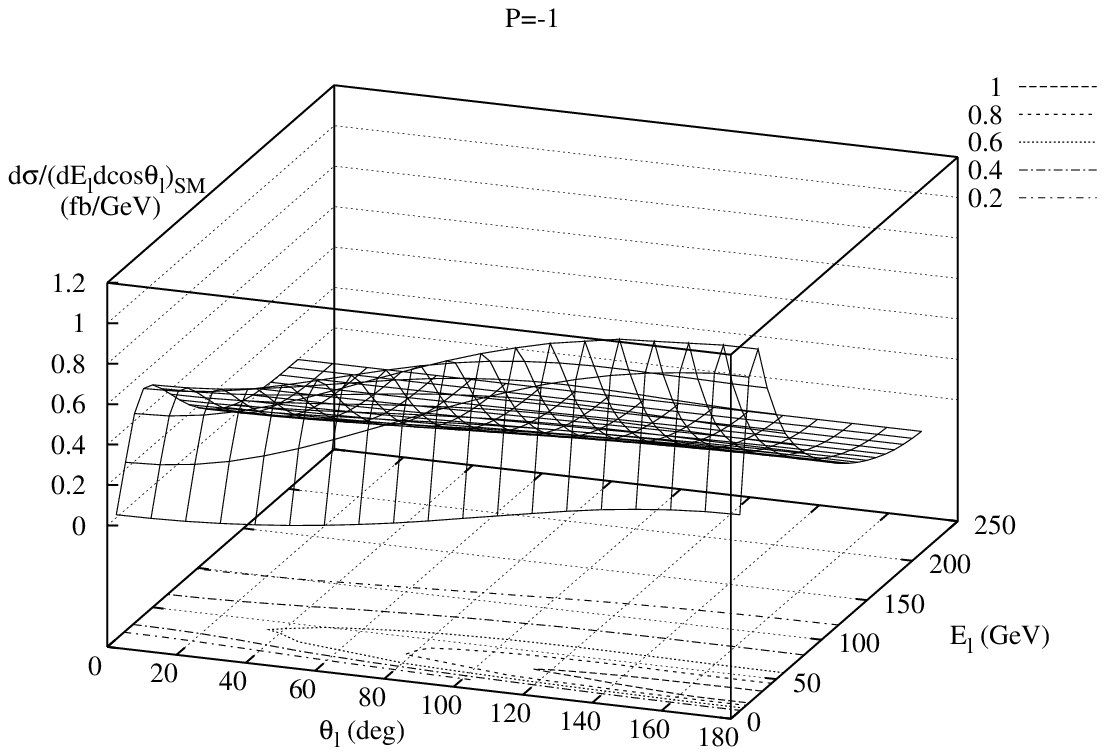,height=7cm }
}}
\caption{The standard model double differential cross section as in Fig.
\ref{distpe0}, but for beam polarizations $P_{e^-}=-P_{e^+}=-1$.  The bottom
plane shows a few contours
of constant cross section.}
\label{distpe-1}
\end{center}
\end{figure}

\begin{figure}[tpb]
\begin{center}
\vskip -2cm
\centerline{\protect\hbox{\psfig{file=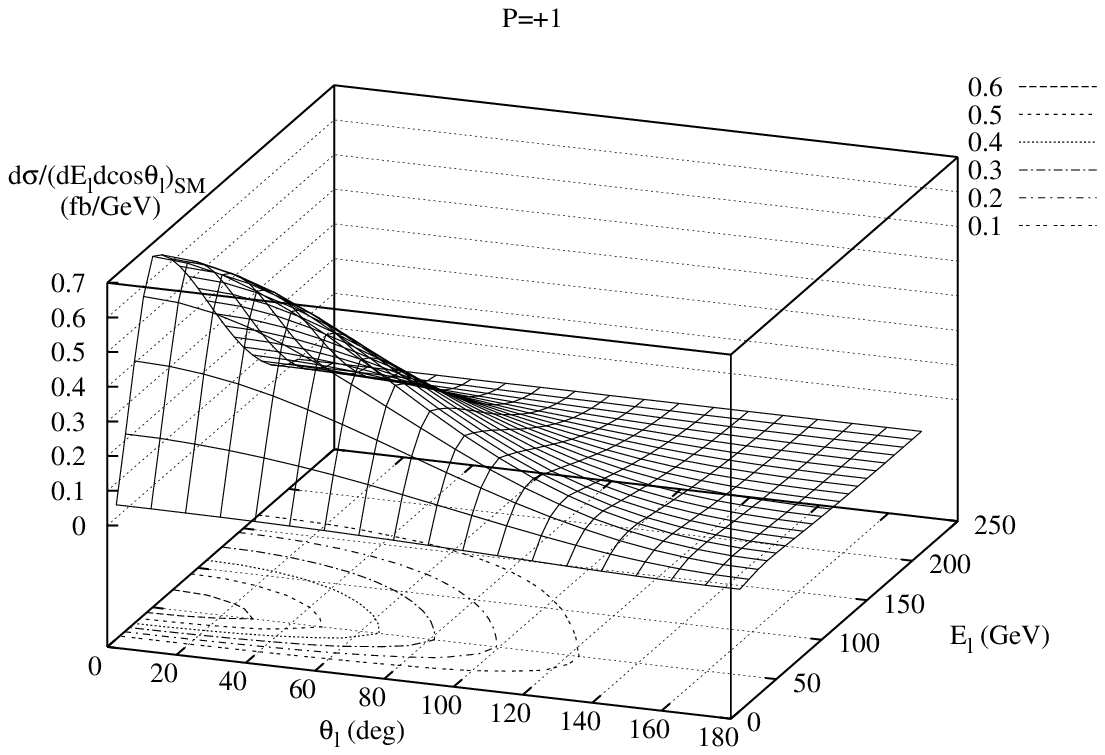,height=7cm}}}
\caption{The standard model double differential cross section as in Fig.
\ref{distpe0}, but for effective beam polarization $P_{e^-}=-P_{e^+}=+1$. The
bottom plane shows a few contours
of constant cross section.}
\label{distpe1}
\end{center}
\end{figure}
 
\begin{figure}[tpb]
\begin{center}
\vskip -2cm
\centerline{\protect\hbox{\psfig{file=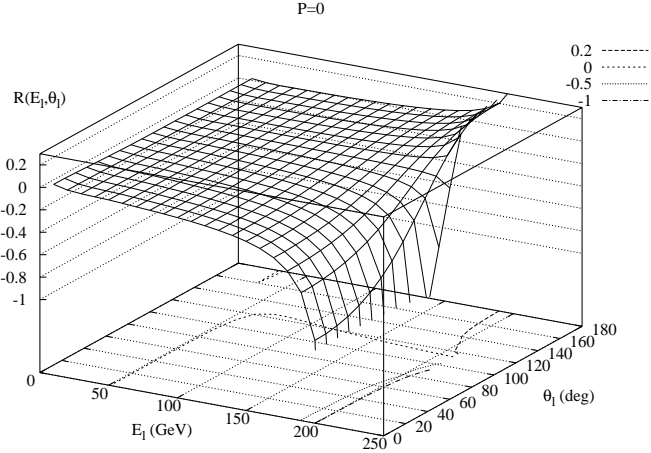,height=7cm}}}
\caption{The ratio $R(E_l,\theta_l)$ of the CP violating part of the cross
section to the SM cross section defined in the text for ${\rm Re}f_{2R}=-{\rm
Re} \o{f}_{2L} = 10^{-2}$ with unpolarized beams. A few contours of constant $R$
 values are shown on the bottom plane.}
\label{distcpvpe0}
\end{center}
\end{figure}

\begin{figure}[tpb]
\begin{center}
\vskip -2cm
\centerline{\protect\hbox{\psfig{file=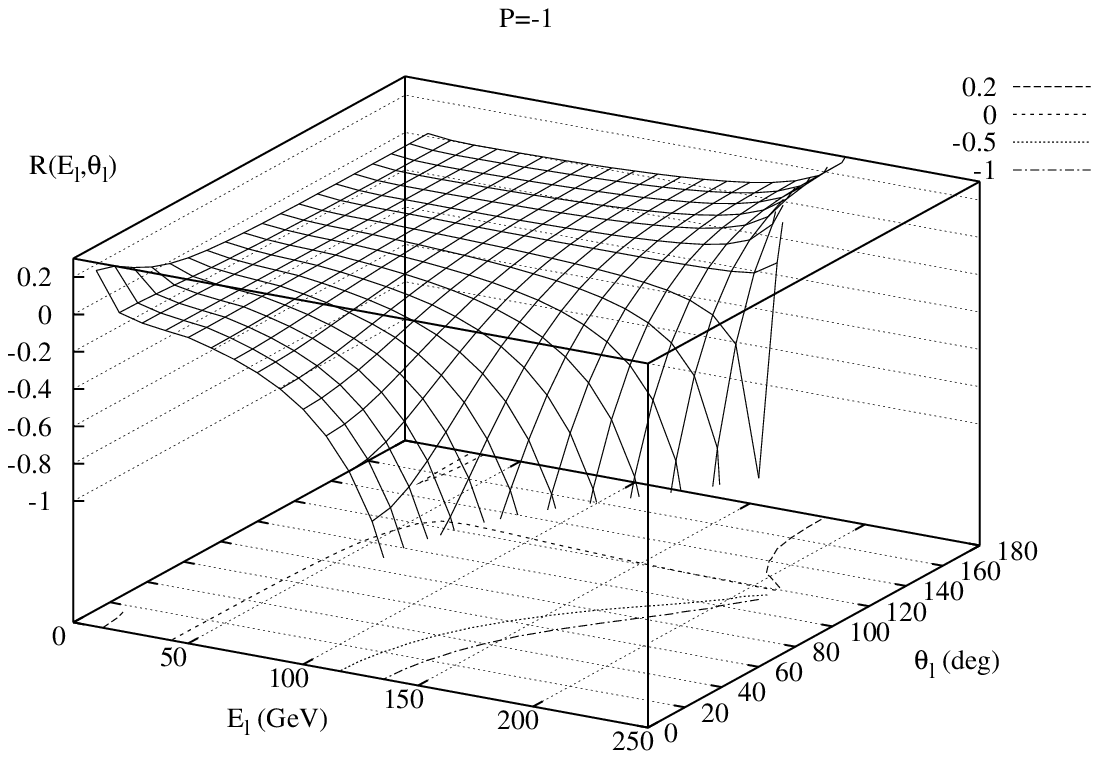,height=7cm}}}
\caption{The ratio $R(E_l,\theta_l)$ of the CP violating part of the cross
section to the SM cross section defined in the text for ${\rm Re}f_{2R}=-{\rm
Re} \o{f}_{2L} = 10^{-2}$ with effective beam polarization $P=-1$. A few
contours of constant values of $R$ are shown on
the bottom plane.}
\label{distcpvpe-1}
\end{center}
\end{figure}

\begin{figure}[tpb]
\begin{center}
\vskip -2cm
\centerline{\protect\hbox{\psfig{file=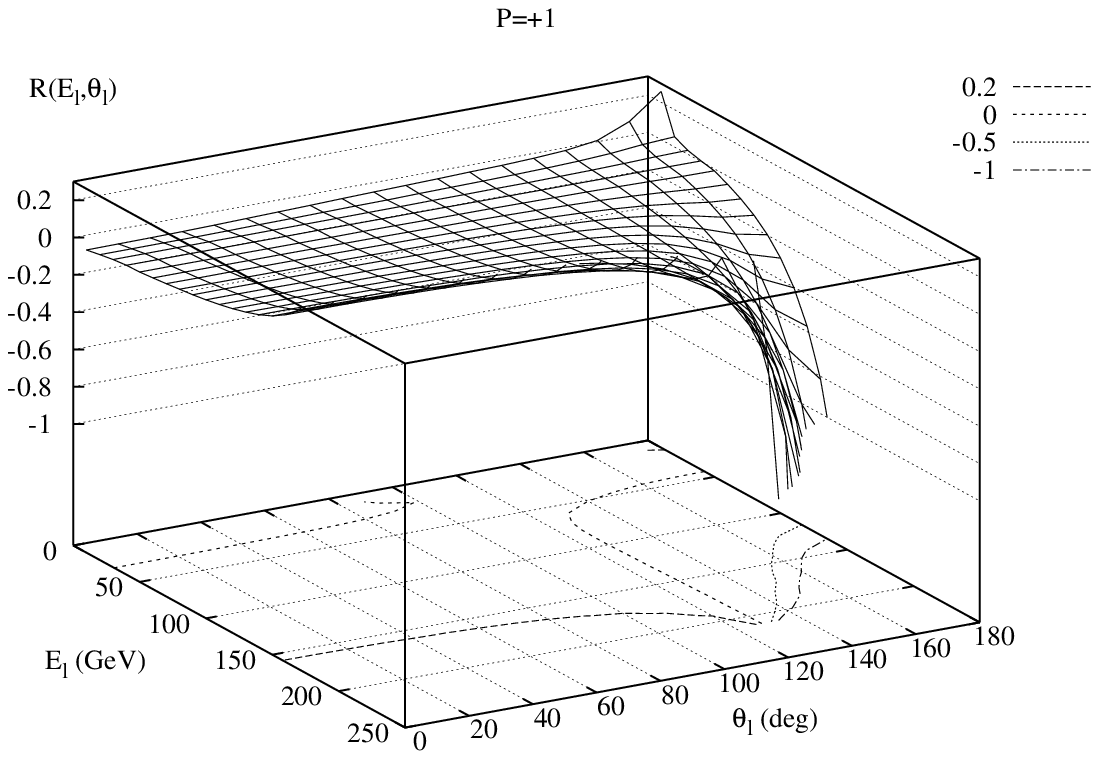,height=7cm}}}
\caption{The ratio $R(E_l,\theta_l)$ of the CP violating part of the cross
section to the SM cross section defined in the text for ${\rm Re}f_{2R}=-{\rm
Re} \o{f}_{2L} = 10^{-2}$ with effective beam polarization $P=+1$. A few
contours of constant values of $R$ are shown on the bottom plane.}
\label{distcpvpe+1}
\end{center}
\end{figure}

In Figs. \ref{distpe0}, \ref{distpe-1} and \ref{distpe1} we give three
dimensional plots of the double differential cross section $d\sigma/(dE_ldc_l)$
as a function of $E_l$ and $c_l$ in SM, for the beam polarizations of
$P_{e^-}=P_{e^+}=0$, 
$P_{e^-}=-P_{e^+}=-1$ and  $P_{e^-}=-P_{e^+}=+1$,
respectively. The plots shown are for
$l^+$. The corresponding ones for $l^-$ are obtained simply by reflecting about
about $\theta_l = \pi/2$, since CP is conserved in SM. In Figs.
\ref{distcpvpe0}, \ref{distcpvpe-1} and \ref{distcpvpe+1} we plot, respectively
for $P=0$, $P=-1$ and $P=+1$, the 
ratio of the CP-violating contribution to the 
differential cross section from only CP violation in the decay to the 
SM value of the differential cross section,
\beq
R(E_l,\theta_l) = \f{
	\displaystyle{\f{d\sigma^{+}}{dE_ldc_l}}(E_l,\theta_l) -
	\displaystyle{	\f{d\sigma^{-}}{dE_ldc_l}}(E_l,\pi-\theta_l)}
{	\displaystyle{	2\f{d\sigma_{\rm SM}}{dE_ldc_l}}(E_l,\theta_l)}
\eeq
 for ${\rm Re}f_{2R}=-{\rm Re} \o{f}_{2L} = 10^{-2}$. In all the
three-dimensional plots, Figs. 1-6, some selected contours are shown on the
bottom plane.

\begin{figure}[ptb]
\begin{center}
\vskip -2cm
\input{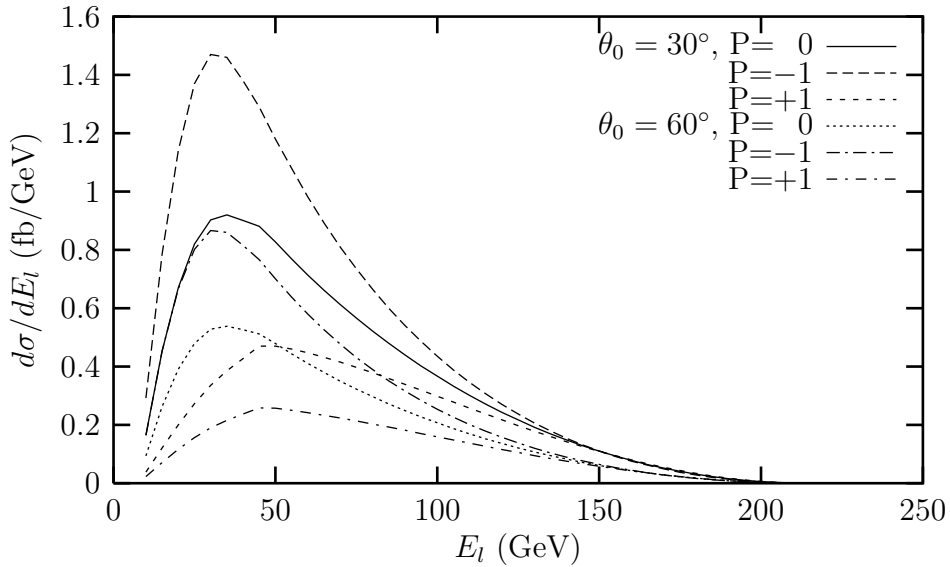}
\caption{SM differential cross section integrated over $\theta_l$
with a cut-off $\theta_0$ in the forward and backward directions.
The curves are shown for $\theta_0=30^{\circ}$ and 
$\theta_0=60^{\circ}$ and for effective beam polarizations $P=0,-1,+1$.}
\label{csplot}
\end{center}
\end{figure}

As noted earlier, in practice there has to be a cut-off in $\theta_l$ in the
forward and backward directions. We shall mainly use a cut-off
$\theta_0=30^{\circ}$, though we have examined the variation of our asymmetries
with cut-off angle in what follows. To get an idea of the effect of cut-off on
event rates, we have plotted the energy dependence of the SM differential cross
section for cut-off values of $\theta_0=30^{\circ}$ and $\theta_0=60^{\circ}$ 
in the forward and backward
directions in Fig. \ref{csplot}, for the polarized and unpolarized cases.

\begin{figure}[tbp]
\begin{center}
\vskip -2cm
\input{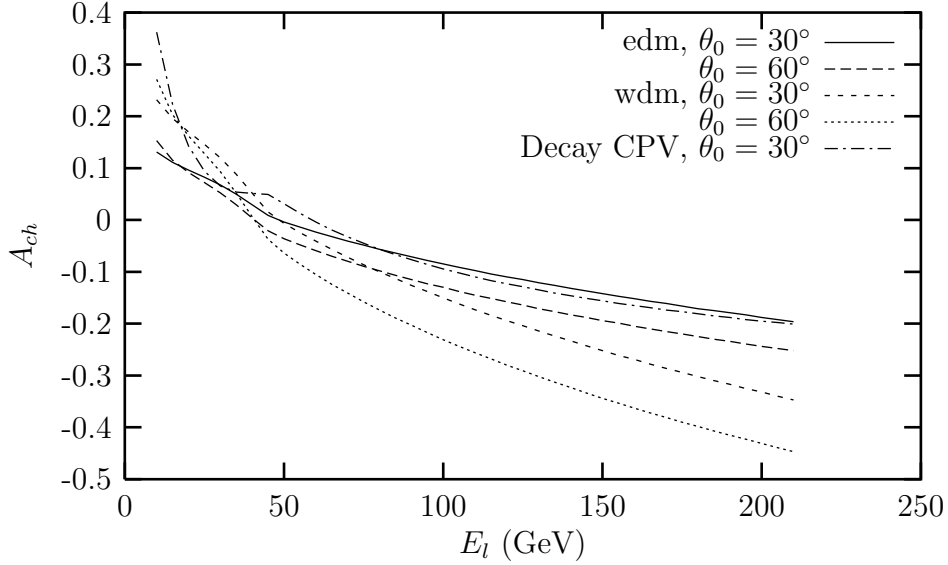}
\caption{The charge asymmetry $A_{ch}$ as function of $E_l$
with two cut-off angles, for the different sources of CP violation,
and unpolarized beams.
The top edm is taken to be $(0.1)\f{e}{m_t}$, the wdm to be
$\f{e}{m_t}$, and ${\rm Re}f_{2R}=-{\rm Re} \o{f}_{2L} = 0.1$.}
\label{chasyunpol}
\end{center}
\end{figure}

In Fig. \ref{chasyunpol}  we plot the lepton charge asymmetry $A_{ch}$
for two values of cut-off angles,
$\theta_0=30^\circ$ and $\theta_0=60^\circ$, for the case of unpolarized $e^+$
and $e^-$ beams. The dependence of the asymmetry on lepton energy is shown
assuming one source of CP violation at a time -- the top electric dipole moment
(edm) $d_t^{\gamma} = (0.1)\f{e}{m_t} \approx 10^{-17}$ $e$ cm, a top weak
dipole moment (wdm) of $d_t^Z = \f{e}{m_t}$, and top decay CP violation (CPV)
corresponding to Re$f_{2R}=-{\rm Re}\o{f}_{2L} = 0.1$. For the last case,
there is no significant change in going from $\theta_0=30^\circ$ to
$\theta_0=60^\circ$. It is interesting to note that all asymmetries change sign
for a lepton energy of 50-60 GeV. As is evident, a value of  0.1 for 
Re$f_{2R}=-{\rm
Re}\o{f}_{2L}$ can produce the same order of asymmetry as top edm
of $10^{-17}$ $e$ cm, or top wdm of $10^{-16}$ $e$ cm.

It should be noted that, as seen from Fig. \ref{csplot},
the SM lepton energy distribution peaks at an energy of
about 40-50 GeV, and falls for low as well as high values of energy. 
Since it is the SM differential cross section which occurs in the denominator
of the charge asymmetry (and the forward-backward asymmetry discussed below),
large values of the asymmetry for low and high values of the lepton energy must
be treated with caution. They can be of order 1 simply because the SM cross
section in the denominator is small for those values of energy.

\begin{figure}[tbp]
\begin{center}
\vskip -2cm
\input{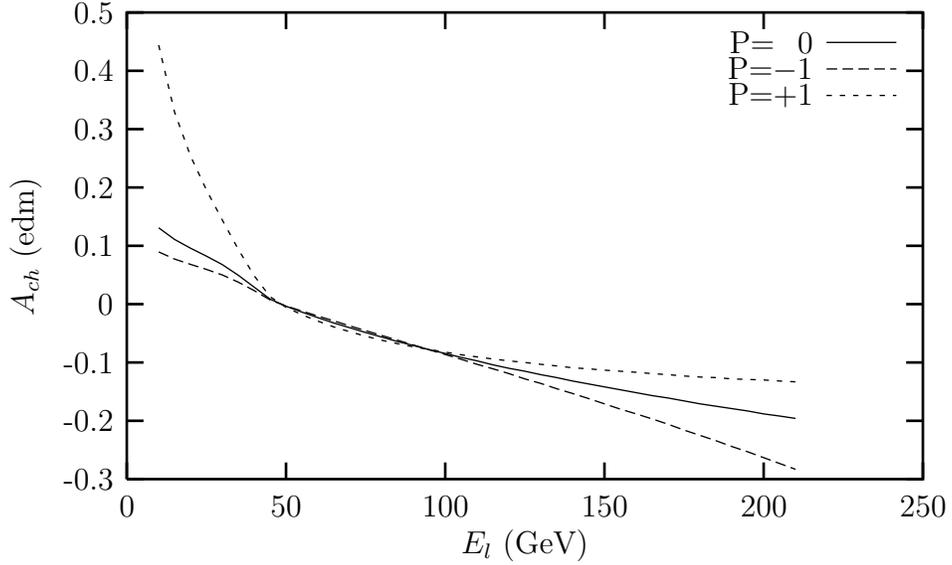}
\caption{The charge asymmetry $A_{ch}$ arising from the top edm
as a function of $E_l$ for $\theta_0=30^{\circ}$ for
three values of the effective $\ee$ polarization $P$. The 
top edm assumed is $(0.1)\f{e}{m_t}$.}
\label{chasyedm}
\end{center}
\end{figure}

\begin{figure}[tbp]
\begin{center}
\vskip -2cm
\input{chasywdm.tex}
\caption{The charge asymmetry $A_{ch}$ arising from the top weak dipole
moment as a function of $E_l$ for $\theta_0=30^{\circ}$ for
three values of the effective $\ee$ polarization $P$. The 
top wdm assumed is $(0.1)\f{e}{m_t}$}
\label{chasywdm}
\end{center}
\end{figure}

\begin{figure}[tbp]
\begin{center}
\vskip -2cm
\input{chasydk.tex}
\caption{The charge asymmetry $A_{ch}$ arising from the CP violating $tbW$
vertex as a function of $E_l$ for $\theta_0=30^{\circ}$ for
three values of the effective $\ee$ polarization $P$. 
Re$f_{2R}=-{\rm Re}\o{f}_{2L} = 0.1$ is assumed.}
\label{chasydk}
\end{center}
\end{figure}

Figs. \ref{chasyedm}, \ref{chasywdm} and \ref{chasydk}
 show the charge asymmetry $A_{ch}$ for a cut-off of
$\theta_0=30^\circ$ for the cases of nonzero edm, wdm, and decay CPV,
respectively, each for effective beam polarization of $P$
of $0$, $-1$ and $+1$.\
It is clear that while charge asymmetries arising from both edm and wdm are
sensitive to polarization for low and high values of lepton energy,
$A_{ch}$ due to decay CPV is not very sensitive to $P$. $A_{ch}$
due to wdm even has opposite signs for $P=+1$ and $-1$. 
This shows that the wdm contribution
to $A_{ch}$ may be easily separated by using data for positive and negative
values of $P$. On the other hand, $A_{ch}$ due to decay CPV may be separated by
concentrating on $E_l\approx 50$ GeV, where the other two asymmetries are close
to zero and change sign. 

\begin{figure}[tbp]
\begin{center}
\vskip -1.5cm
\input{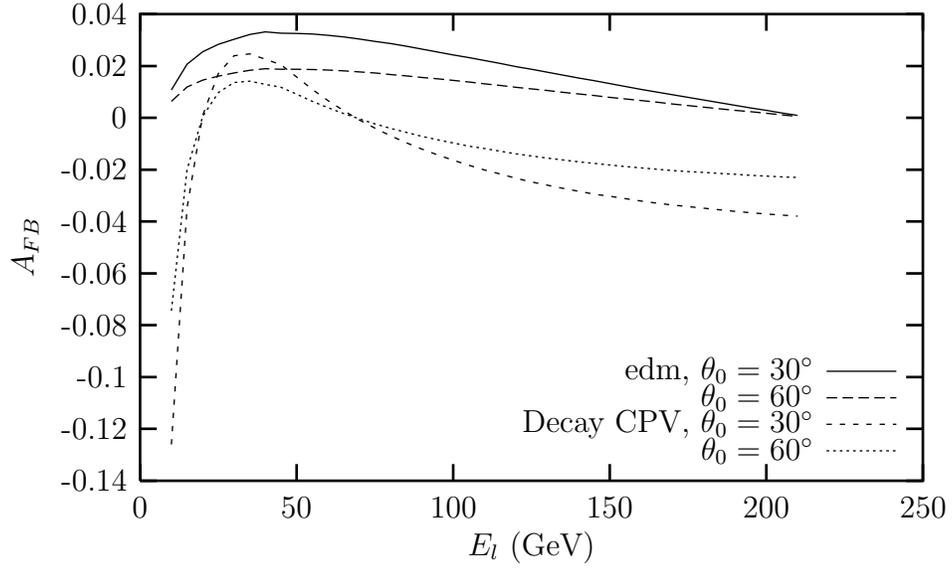}
\caption{The asymmetry $A_{FB}$ as function of $E_l$
with two cut-off angles, for the different sources of CP violation,
for unpolarized beams.
The top edm is taken as $(0.1)\f{e}{m_t}$,
 and ${\rm Re}f_{2R}=-{\rm Re} \o{f}_{2L} = 0.1$.
The curve for a top wdm of $\f{e}{m_t}$ 
approximately coincides with the curve for edm, and is not shown separately.}
\label{fbasyunpol}
\end{center}
\end{figure}

Fig. \ref{fbasyunpol}, which is the analogue of Fig. \ref{chasyunpol}, 
shows the forward-backward asymmetry
(symmetrized over lepton charges) $A_{FB}$. It is seen from the figure that the
edm contribution dominates over the wdm contribution for equal values
of $d_t^{\gamma}$ and $d_t^Z$, and in fact, it is very nearly 10 times the wdm
contribution, by accident.
Unlike $A_{ch}$, there is no change of sign.
$A_{FB}$ due to decay CPV, on the other hand, changes sign twice. Now the
asymmetries are larger for a cut-off $\theta =30^{\circ}$ rather than for
$\theta =60^{\circ}$, unlike in the case of $A_{ch}$.

\begin{figure}[tbp]
\begin{center}
\vskip -1.7cm
\input{fbasyedm.tex}
\caption{The asymmetry $A_{FB}$ arising from the top electric dipole
moment as a function of $E_l$ for $\theta_0=30^{\circ}$ for
three values of the effective $\ee$ polarization $P$. The 
top edm assumed is $\f{e}{m_t}$.}
\label{fbasyedm}
\end{center}
\end{figure}

\begin{figure}[tbp]
\begin{center}
\vskip -1.7cm
\input{fbasywdm.tex}
\caption{The asymmetry $A_{FB}$ arising from the top weak dipole
moment as a function of $E_l$ for $\theta_0=30^{\circ}$ for
three values of the effective $\ee$ polarization $P$. The 
top edm assumed is $\f{e}{m_t}$.}
\label{fbasywdm}
\end{center}
\end{figure}

\begin{figure}[tbp]
\begin{center}
\vskip -1.7cm
\input{fbasydk.tex}
\caption{The asymmetry $A_{FB}$ arising from the CP violating $tbW$
vertex as a function of $E_l$ for $\theta_0=30^{\circ}$ for
three values of the effective $\ee$ polarization $P$. 
Re$f_{2R}=-{\rm Re}\o{f}_{2L} = 0.1$ is assumed.}
\label{fbasydk}
\end{center}
\end{figure}

Looking at the polarization dependences of $A_{FB}$ in Figs. \ref{fbasyedm},
\ref{fbasywdm} and \ref{fbasydk} it is clear
that in all cases there is strong sensitivity to $P$. Again $A_{FB}$ due to wdm
changes sign with $P$, whereas that due to edm does not. However, $A_{FB}$ due
to decay CPV also changes sign with $P$ for low as well as high values of
$E_l$.

It is useful to get an idea of the sensitivity of a linear collider with an
integrated luminosity of 100 fb$^{-1}$ to the CP-violating parameters. The
number of events expected in an energy bin of 10 GeV centred around $E_l\approx
40$ GeV in the unpolarized case 
would be $N\approx 900$, corresponding to $\sqrt{N}=30$. For a
$2\sigma$ effect, the asymmetry then should be at least $0.06$. A charge
asymmetry of this magnitude would correspond to edm of about $10^{-18}$ $e$ cm,
or wdm of about $10^{-18}$ $e$ cm, or $f^{\pm}\approx 0.06$.
Combining data from several bins and doing a likelihood analysis 
could easily improve this sensitivity to $f^{\pm}
\approx 10^{-3}$. Since the expectation from popular models like the minimal
supersymmetric standard model for $f^{\pm}$ is in this range, it would be
interesting to look for the asymmetries we have discussed. A more detailed
study of the statistical significance and possible backgrounds 
would be worthwhile.

To summarize, after obtaining analytic expressions for angular distributions
and energy-angle double distributions including anomalous effects in production
as well as decay, we have studied studied the CP-violating asymmetries $A_{ch}$
and $A_{FB}$ as functions of decay-lepton energy and the initial beam
polarization. Since anomalous effects in the decay do not appear in the angular
distributions where energy has been integrated over, these are not useful for
the study of CP violation in decay. On the other hand, these are most useful
for the study of CP violation in production, as discussed in \cite{ppasymm}. 

To be able to get a handle on CP violation in the $tbW$ vertex, we have
compared the $E_l$ and polarization dependence of $A_{ch}$ and $A_{FB}$ arising
from the top-quark edm, wdm, and decay CPV separately. We find interesting
features like zeros and sign changes in the asymmetries as a function of
energy, which are different for the different sources of CP violation. In
general there is strong dependence on effective beam polarization $P$,
which not only enhances the asymmetry in most cases, but might 
also help in discrimination amongst the various sources of CP violation. We
have not 
studied in detail the procedure for discriminating, but have indicated
significant features which might be used. A detailed study of various
sensitivities would be useful.

We end with a few comments.

The CP-violating asymmetries we have considered are simple in principle as also
straightfoward to implement from the experimental point of view, as they do not
require the determination of the top-quark direction or momentum. It is
possible to use correlations of optimal observables, which would maximize the
statistical sensitivity \cite{atwood}, at the expense of simplicity.

Note that the asymmetries we have chosen are odd under CP, but even under
``naive" time reversal T$_{\rm N}$, i.e., sign reversal of spins and momenta,
without and interchange of initial and final states. The CPT theorem therefore
implies that they should necessarily come from the absorptive part of the
amplitude \cite{whepp}. As noted earlier, only the absorptive part of the $tbW$
vertex contributes to the differential cross sections. It follows that had a
CP-odd asymmetry which was odd under T$_{\rm N}$ been chosen, it could not have
depended on CP violation in decay. 

In the above, we have taken into account the practical requirement of imposing
a cut on the angle of the detected lepton with respect to the beam axis. A
similar practical constraint might be necessary for the lepton energy. The
detection of a charged lepton will require it to have a minimum energy.
However, for a cm energy of 500 GeV, the minimum lepton energy allowed
kinematically is about 7.5 GeV. Thus, if the cut needed for detection is not
required to be below about 7.5 GeV, our results can be used as such. If,
however, a cut is to be larger, this cut may itself introduce a dependence on
$f^{\pm}$ of the polar angle distribution, which was found to be absent when the
full range of energy is integrated over, as in eq. (\ref{cldist}). This would
need further study. However, it can be seen that this question can be easily
handled by doing an analytic integration over the appropriate energy region
using our expressions.

After the completion of this work, the paper of E. Boos {\it et al.} \cite{boos}
was listed
in the Los Alamos archive, which discusses asymmetries due to anomalous $tbW$
vertex. They do not discuss explicitly CP-violating asymmetries which have been
described in this paper.

{\bf Acknowledements} I thank Zenro Hioki for encouragement and correspondence,
and also for providing independent confirmation of the result in eq.
(\ref{cldist}) prior to publication. I thank Debajyoti Choudhury for a
clarification on energy cuts for charged leptons. I also thank Toni Pich for
his suggestion that beam polarization might be able to discriminate CP
violation in decay from CP violation in production.

\end{document}